# Towards Energy-Efficient Database Cluster Design

Willis Lang
University of Wisconsin
wlang@cs.wisc.edu

Stavros Harizopoulos
Nou Data
stavros@noudata.com

Jignesh M. Patel
University of Wisconsin
jignesh@cs.wisc.edu

Mehul A. Shah
Nou Data
mehul@noudata.com

Dimitris Tsirogiannis
Microsoft Corp.
dimitsir@microsoft.com

## ABSTRACT

Energy is a growing component of the operational cost for many "big data" deployments, and hence has become increasingly important for practitioners of large-scale data analysis who require scale-out clusters or parallel DBMS appliances. Although a number of recent studies have investigated the energy efficiency of DBMSs, none of these studies have looked at the architectural design space of energy-efficient parallel DBMS clusters. There are many challenges to increasing the energy efficiency of a DBMS cluster, including dealing with the inherent scaling inefficiency of parallel data processing, and choosing the appropriate energy-efficient hardware. In this paper, we experimentally examine and analyze a number of key parameters related to these challenges for designing energy-efficient database clusters. We explore the cluster design space using empirical results and propose a model that considers the key bottlenecks to energy efficiency in a parallel DBMS. This paper represents a key first step in designing energy-efficient database clusters, which is increasingly important given the trend toward parallel database appliances.

## 1. INTRODUCTION

In recent years, energy efficiency has become an important database research topic since the cost of powering clusters is a big component of the total operational cost [13, 15, 19]. As "big data" becomes the norm in various industries, the use of clusters to analyze ever-increasing volumes of data will continue to increase. In turn, this trend will drive up the need for designing energy-efficient data processing clusters. The focus of this paper is on designing such energy-efficient clusters for database analytic query processing.

One important problem regarding the energy efficiency of database clusters surrounds the classical problem of non-linear scalability in parallel data processing [12]. Non-linear scalability refers to the inability of a parallel system to proportionally increase performance as the computational resources/nodes are increased. More recently, it was shown that modern parallel DBMSs are still subject to these scalability pitfalls that affect their performance and increase their total operating cost [25, 31].

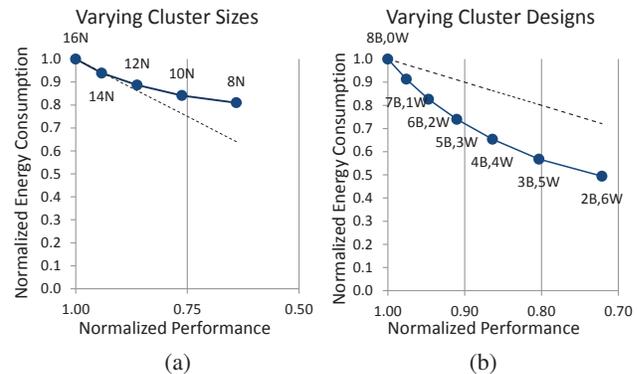

**Figure 1: (a) Empirical energy consumption and performance results for Vertica running TPC-H Q12 (at scale factor of 1000). The dotted line indicates trading an X% decrease in performance for an X% decrease in energy consumed so that the Energy Delay Product (EDP= $energy \times delay$) metric is constant relative to the 16 node cluster. (b) Modeled performance and energy efficiency of an 8 node cluster made of various traditional Beefy nodes and low-power Wimpy nodes when performing a parallel hash join on our custom parallel execution engine P-store. Wimpy nodes only scan and filter the data before shuffling it to the Beefy nodes for further processing. A constant EDP relative to the all Beefy cluster is shown by the dotted line.**

### 1.1 Illustrative Experiments

First, we present two results that show how varying the cluster size and changing the cluster design can allow us to trade performance for reduced energy consumption for a single query workload.

In this paper, we studied the effect of this undesirable scalability phenomenon on the energy efficiency of parallel data processing clusters using a number of parallel DBMSs, including Vertica and HadoopDB. Our first result is shown in Figure 1(a) for TPC-H Q12 (at scale factor 1000) for Vertica. In this figure, we show the relative change, compared to a 16 node cluster, in the energy consumed by the cluster and the query performance[1] as we decrease the cluster size two nodes at a time (please see Section 3.1 for more details). Next, we discuss three key insights that can be drawn from Figure 1(a), which also shape the theme of this paper.

First, this result shows the classic sub-linear parallel speedup phenomenon; namely, given a fixed problem size, increasing the computing resources by a factor of $X$ provides *less than* an $X$ times increase in performance [12]. Or, conversely, decreasing the resources to $1/X$, results in a relative performance *greater than* $1/X$. We can observe this phenomenon in Figure 1(a) because reducing the cluster size from 16 nodes (**16N**) to eight nodes (**8N**), results in a performance ratio greater than 50%. Note that since performance is the inverse of the response time, a performance ratio greater than 50% at 8N means that the response time at 16N is more than half the response time at 8N (i.e., sub-linear speedup).

---
[1]Here performance is the inverse of the query response time.

Second, as shown by the solid curve in Figure 1(a), the total energy required to execute the query decreases as we reduce the cluster size from 16N, even though it takes longer to run the query. Recall that energy is a product of the average power drawn by the cluster and the response time of the query. Due to the sub-linear speedup, going from 16N to 8N reduces the performance by only 36%, but the average power drops by roughly half (since we have half the number of nodes). Consequently, the energy consumption ratio (relative to the 16N case) for fewer than 16 nodes is *less than 1.0*. This is an encouraging energy efficiency gain, albeit at the cost of increased query response time.

Lastly, in Figure 1(a), the dotted line shows the line where the *Energy Delay Product* (EDP) is constant, relative to the 16N case. EDP is defined as $energy \times delay$ (measured in Joules seconds), and is commonly used in the architecture community as a way of studying designs that trade-off energy for performance[2]. Here "energy" refers to the query energy consumption and "delay" refers to the query response time. A constant EDP means that we have traded $x\%$ of performance for an $x\%$ drop in energy consumption. Preferably, it would be nice to identify design points that lie below the EDP curve, as such points represent trading proportionally less performance for greater energy savings.

In Figure 1(a), all the actual data/design points (on the solid line) are above the EDP curve. In other words, as we reduce the cluster size from 16 nodes to 8 nodes, we are giving up proportionately more performance than we are gaining in energy efficiency. For example, the 10 node configuration (**10N**) pays a 24% penalty in performance for a 16% decrease in energy consumption over the 16N case. Such trade-offs may or may not be reasonable depending on the tolerance for performance penalties, but the EDP curve makes it clear in which directions the trade-offs are skewed. This observation motivates the key question that is addressed in this paper: *What are the key factors to consider when we design an energy-efficient DBMS cluster so that we can favorably trade less performance for more energy savings (i.e., lie below the EDP curve)?*

To understand the reasons why our observed data points lie above the EDP curve in Figure 1(a), and to carefully study the conditions that can produce design points that lie below the EDP curve, we built and modeled a custom parallel data execution kernel called P-store. The data that we collected from real parallel DBMSs was used to validate the performance and energy model of P-store (see Section 4 for details). Then, using P-store, we systematically explored both the query parameters and the cluster design space parameters to examine their impact on performance and energy efficiency. We also developed an analytical model that allows us to further explore the design space.

One of the interesting insights we found using both P-store and our analytical model is that there are query scenarios where certain design points lie below the EDP curve. One such result is shown in Figure 1(b). Here, we use our analytical model to show the energy versus performance trade-off for various eight node cluster designs, when performing a join between the TPC-H LINEITEM and the ORDERS tables (see Section 5 for details). In this figure, similar to Figure 1(a), we plot the relative energy consumed by the system and the response time against a reference point of an eight node Xeon-based ("**B**eefy") cluster. We then gradually replaced these nodes with mobile Intel i7 based laptops ("**W**impy")[3] nodes. The Wimpy nodes do not have enough memory to build in-memory hash tables for the join, and so they only scan and filter the table data before shuffling them off to the Beefy nodes (where the actual join is performed). As opposed to Figure 1(a), which is an experiment done with homogeneous Beefy nodes, Figure 1(b) shows data points below the EDP curve. This result is interesting as it shows that it is possible to achieve a relatively greater energy savings than response time penalty (to lower the EDP) when considering alternative cluster designs.

---
[2]With the growing viewpoint of considering an entire cluster as a single computer [10], EDP is also a useful way of thinking about the interactions between energy consumption and performance when designing data centers [22, 38].

Energy-efficient cluster design with potentially heterogeneous cluster nodes needs to be considered since non-traditional heterogeneous clusters are now showing up as database appliances, such as Oracle's Exadata Database Machine [1]. Thus, to a certain extent, commercial systems designers have already started down the road to heterogeneous clusters and appliances. Such considerations may also become important if future hardware (e.g., processor and/or memory subsystems) allows systems to dynamically control their power/performance trade-offs. Our paper provides a systematic study of both the software and the hardware design space parameters in an effort to draw conclusions about important design decisions when designing energy-efficient database clusters.

## 1.2 Contributions

The key contributions of this paper are: first, we empirically examine the interactions between the scalability of parallel DBMSs and energy efficiency. Using three DBMSs: Vertica, HadoopDB, and a custom-built parallel data processing kernel, we explore the trade-offs in performance versus energy efficiency when performing speedup experiments on TPC-H queries (e.g., Figure 1(a)). From these results, we identify distinct bottlenecks for performance and energy efficiency that should be avoided for energy-efficient cluster design.

Second, non-traditional/wimpy low-power server hardware has been evaluated for its performance/energy-efficiency trade-offs, and we leverage these insights along with our own energy-efficiency micro-benchmarks to explore the design space for parallel DBMS clusters. Using P-store, we provide a model that takes into account these different server configurations and the parallel data processing bottlenecks, and predicts data processing performance and energy efficiency for various node configurations of a database cluster.

Third, using our model we study the design space for parallel DBMS clusters, and illustrate interesting cluster design points, under varying query parameters for a typical hash join (e.g., Figure 1(b)).

Finally, we organize the insights from this study as (initial) guiding principles for energy-efficient data processing cluster design.

The remainder of this paper is organized as follows: Section 2 contains background discussion and related work; Section 3 discusses our findings with Vertica and HadoopDB. In Section 4 we present P-store. In Section 5, we model P-store and use the model to explore interesting cluster design points. Section 6 summarizes our findings as guiding principles for energy-efficient data processing cluster design. We make our concluding remarks in Section 7.

## 2. BACKGROUND AND RELATED WORK

With recent improvements in power delivery, cooling and heat extraction, significant improvements have been made in improving the energy efficiency of large data centers [13]. At the server level, there are two ways to improve efficiency. One is to consolidate work onto few servers and turn off unused servers [23, 24, 27]. However, switching servers on and off has direct costs such as increased query latency and decreased hardware reliability. Another approach is to consolidate the server use for a given task, and improve the scheduling and the physical design to use the remaining servers for other tasks. In this paper, we tackle the latter issue to explore energy-efficient cluster design.

---
[3]We use the term "wimpy" as in [37], i.e. "slower but [energy] efficient".

| DBMS | Vertica | | RAM | 48GB |
|---|---|---|---|---|
| # nodes | 16 | | Disks | 8x300GB |
| TPC-H size | 1TB (scale 1000) | | Network | 1Gb/s |
| CPU | Intel X5550 2 sockets | | SysPower | $130.03C^{0.2369}$ |
| | | | | C = CPU utilization |

Table 1: Cluster-V Configuration

There has been a growing number of efforts to improve the energy efficiency of clusters [7–10, 20, 23–25, 27, 32, 34, 36]. We have seen holistic redesigns that treat a data center as a single computer [9, 10, 32], and approaches that consider workload consolidation techniques in order to meet the power constraints and reduce the energy requirements of clusters [7, 23, 24, 27, 34]. In data analytics environments, delaying execution of workloads (increasing response time) due to energy concerns have been proposed [20, 23].

A number of studies have also considered the use of low-power ("wimpy") nodes consisting of low-power storage (SSDs) and processors (mobile CPUs) [8, 17, 25, 26, 33, 36]. Primarily, these designs target computationally "simple" data processing tasks that are partitionable, such as key-value workloads [37]. For such workloads, wimpy node clusters are more energy efficient compared to traditional clusters built using more power-hungry "beefy" nodes. But, for analytical database workloads that often exhibit non-linear speedup, and especially when the network is a bottleneck, it was shown that such wimpy node clusters often increase the total operating costs [25]. Alternatively, servers can be augmented with custom, low-power circuitry, like Field-Programmable Gate Arrays [2, 29]. Such approaches are complementary to our study of heterogeneous cluster design, and are an interesting direction for future work.

Early database research studies speculated that the database software has an important role to play in improving the energy efficiency, and argued for the redesign of several key components such as the query optimizer, the workload manager, the scheduler and the physical database design [14, 18, 22]. Many of these suggestions assumed that, like cars, computer systems have different optimal performance and energy efficiency points. However, only preliminary experimental data was provided to support these claims.

Subsequent efforts have studied how alternate database designs and configurations can improve the energy efficiency of a DBMS [21, 35, 38]. Meza et al. [28], have studied the energy efficiency of a scale-up enterprise decision support system, and showed that the most energy-efficient configuration was not always the fastest depending on the degree of declustering across disks. Lang et al. [24], showed how data replication can be leveraged to reduce the number of online cluster nodes in a parallel DBMS. That work is complimentary to ours as we could leverage similar replication techniques to dynamically augment cluster size. Chen et al. [11], considered heterogeneous usage of MapReduce servers through a software workload manager. In contrast, we consider heterogeneity in the cluster servers as well as the execution paths across servers.

## 3. PARALLEL DATABASE BEHAVIOR

In this section we examine the performance behavior and speedup in shared-nothing DBMS clusters, and examine its effect on energy efficiency. We see that bottlenecks that degrade performance, like network bottlenecks, decrease the energy efficiency of a cluster design. We experiment with two off-the-shelf, column-oriented, parallel DBMSs, Vertica [3] and HadoopDB [6] (with Vectorwise [4]). Vertica was deployed on *cluster-V* (see Table 1), and we used queries from the TPC-H benchmark at scale factor 1000.

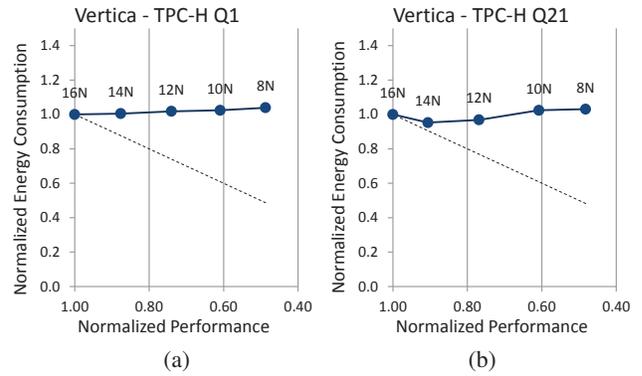

Figure 2: Vertica TPC-H (a) Q1, (b) Q21 (scale 1000) empirical parallel speedup results and its effect on energy efficiency under various cluster sizes. Cluster details can be found in Table 1. The dotted line indicates trading a proportional decrease in performance for a decrease in energy consumed such that the EDP ($= energy \times delay$) metric is constant.

### 3.1 Vertica

We used Vertica v.4.0.12-0 64 bit RHEL5 running on 16 HP ProLiant DL360G6 servers (the cluster configuration is described in Table 1), varying the cluster size between 8 and 16 nodes, in 2 node increments. We only present results with a warm buffer pool. Given the use of column-store in Vertica, the working sets for all the queries fit in main memory (even in the 8 node case).

We did not have physical access to our clusters so we used real power readings from the iLO2 remote management interface [5] to develop server power models based on CPU utilization, following an approach that has been used before [34]. Using a single cluster-V node, we used a custom parallel hash-join program (see Section 4) to generate CPU load, and iLO2 measured the reported power drawn by the node usage. We varied the number of concurrent joins to control the utilization on the nodes. iLO2 reports measurements averaged over a 5 minute window, and we ran the experiments for three 5 minute windows for each CPU utilization measurement. The power readings at each CPU utilization level were stable, and we use the average of the three readings to create our server power models. (Our measurements of physically accessible servers, not using iLO2, in Section 5 produced similar models.) In the interest of space, we omit the measured results but show the derived power models in Table 1 as "SysPower". (In this paper, we explored exponential, power, and logarithmic regression models, and picked the one with the best $R^2$ value.)

We employed Vertica's hash segmentation capability which hash partitions a table across the cluster nodes on a user-defined attribute. We partitioned the LINEITEM, ORDERS, and CUSTOMER tables using the hash segmentation, while all the remaining TPC-H tables were replicated on each node. The ORDERS and the CUSTOMER tables were hashed on the O_CUSTKEY and C_CUSTKEY attributes respectively, so that a join between these two tables does not require any shuffling/partitioning of the input tables on-the-fly. The LINEITEM table was hashed on the L_ORDERKEY attribute.

We ran a number of TPC-H queries and present a selection of our results in this section. In Figure 2(a) we show the energy consumed and performance (i.e., the inverse of the query response time) for various cluster sizes running the TPC-H Query 1. This query does

1686

not involve any joins and only does simple aggregations on the LINEITEM table. The data points are shown relative to the largest cluster size of 16 nodes, and the "break-even" EDP line is also plotted, as was also done earlier in Figure 1(a). Recall from Section 1 that this dotted line represents data points that trade energy savings for a performance penalty such that the EDP remains constant.

In Figure 2(a), we observe that Vertica's performance scales linearly, as the 8 node cluster has a performance ratio of about 0.5 compared to the 16 node case. Consequently, the energy consumption ratio is fairly constant since the 50% performance degradation is offset by a 50% drop in average cluster power. This result is important because it shows that a partitionable analytics workload (like TPC-H Query 1), exhibits ideal speedup as we allocate more nodes to the cluster. Thus the energy consumption will remain roughly constant as we change the cluster size. In other words, one interesting energy-efficient design point is to simply provision as many nodes as possible for this type of query (as there is no change in energy consumption, but there is a performance penalty).

Let us consider a more complex query. Figure 2(b) shows the results for TPC-H Query 21, which is a query that involves a join across four tables: SUPPLIER, LINEITEM, ORDERS, and NATION. The SUPPLIER and NATION tables were replicated across all the nodes, so only the join between the LINEITEM and the ORDERS tables on the ORDERKEY attribute required repartitioning (of the ORDERS table on O_ORDERKEY). Besides the four table join, Query 21 also contains SELECT subqueries on the LINEITEM table within the WHERE clause.

Surprisingly, the results for the more complex TPC-H Query 21 results, shown in Figure 2(b), is similar to that of the simpler TPC-H Query 1, which is shown in Figure 2(a). Since both queries scale well, the energy consumption is fairly flat in both cases. It is interesting to consider why the more complex TPC-H Query 21 scales well, even though it requires a repartitioning of the ORDERS table during query processing. The reason for this behavior is that the bulk of this query (94.5% of the total query time for eight nodes – **8N**) is spent doing node local execution. Only 5.5% of the total query time is spent repartitioning for the LINEITEM and ORDERS join. Since the bulk of the query is processed locally on each node, this means that increasing the cluster size increases performance nearly linearly. Thus, for a query with few bottlenecks, Vertica exhibits nearly ideal speedup, and so the energy-efficient cluster design is to simply use as many nodes as possible.

However, we have already seen a complex query where Vertica *does not* exhibit ideal speedup. Compare the TPC-H Query 21 result to Query 12 (shown in Section 1, Figure 1(a)), which is a much simpler two table join between the ORDERS and the LINEITEM tables, and performs the same repartitioning as Query 21. Compared to the 5.5% time spent network bottlenecked during repartitioning in Query 21, Query 12 spends 48% of the query time network bottlenecked during repartitioning with the eight node cluster. Since the proportional amount of total query time spent doing node local processing is now dramatically reduced, we see in Figure 1(a), that increasing the cluster size does not result in a proportional increase in performance. As such, the energy efficiency suffers dramatically as we increase the cluster size from 8 to 16 nodes.

**Summary:** Our Vertica results show that for queries that do not involve significant time repartitioning (i.e., most of the query execution time is spent on local computation at each node), the energy consumption vs. performance curves are roughly flat. This implies that an interesting point for energy-efficient cluster design is to pick a cluster that is as large as possible, as there is no energy savings when using fewer nodes, but there is an increase in query response time. However, for queries that are bottlenecked, as we saw with TPC-H Q12's network repartitioning, non-linear speedup means that a potential energy-efficient design decision is to reduce the cluster size up to the point where the lower performance is acceptable[4]. Of course, one simple way to mitigate repartitioning bottlenecks is to devise energy-aware repartitioning or *replication* strategies. An analysis and comparison to such strategies is beyond the scope of this paper, but an interesting target of future work.

### 3.2 HadoopDB

HadoopDB is an open-source, shared-nothing parallel DBMS that uses Hadoop to coordinate independent nodes, with each node running Ingres/VectorWise [4, 6]. We used Hadoop ver. 0.19.2 and Ingres VectorWise ver. 1.0.0-112. Setup scripts for HadoopDB and the TPC-H queries we ran were provided by the authors of [6].

In HadoopDB, Hadoop acts as a means of communication between the individual nodes. However, Hadoop was designed with fault tolerance as one of the primary goals and consequently, the performance of our version of HadoopDB was limited by the Hadoop bottleneck. Our evaluation of HadoopDB found that (similar to the results in Figure 1(a)) the best performing cluster is not always the most energy-efficient. In the interest of space, we omit these results.

### 3.3 Discussion

From our results with Vertica, we have found that there are queries where the highest performing cluster configuration is *not* the most energy efficient (TPC-H Q12, Figure 1(a)) due to a network bottleneck. With a commercial DBMS, simpler queries that do not require any internode communication scale fairly linearly with increased cluster size. We saw this with Query 1 and Query 21 (Figures 2(a,b)). However, with communication heavy queries such as Q12 (Figure 1(a)), increasing the cluster size simply adds extra network overhead that dampens the performance increase and reduces our energy savings. *Thus, we conclude that optimizing for parallel DBMS performance does not always result in the lowest energy consumed per query and this is the opposite conclusion from prior work which dealt with single server DBMS environments [35].*

With our results, we hypothesize that queries with bottlenecks, such as the network or disk, cause node underutilization and thus, the bottlenecks decrease the energy efficiency as the cluster size increases. Since Vertica and HadoopDB are black-box systems, to further explore this hypothesis of performance bottlenecks affecting affecting energy efficiency, we built a custom parallel query execution engine called P-store. Using P-store we empirically and analytically studied various join operations that require varying degrees of network repartitioning.

## 4. ENERGY EFFICIENCY OF A PARALLEL DATA PROCESSING ENGINE

From our empirical, off-the-shelf, parallel DBMS observations, we concluded that the scalability of the system running a given query plays an important role in influencing the energy efficiency of the cluster design points. Typically, poor scalability is a consequence of bottlenecks in the system. We now describe some of these bottlenecks, and then present our custom parallel query execution engine P-store that allows us to study these bottlenecks in more detail.

### 4.1 Bottlenecks

Since there can be a multitude of implementation-specific issues that can affect DBMS scalability, in this paper we are interested only in fundamental bottlenecks that are inherent to the core engine.

---
[4]Cluster systems often have implicit or explicit minimum performance targets for many workloads. We recognize that determining when such limits are acceptable is a broad and emerging research topic.

1687

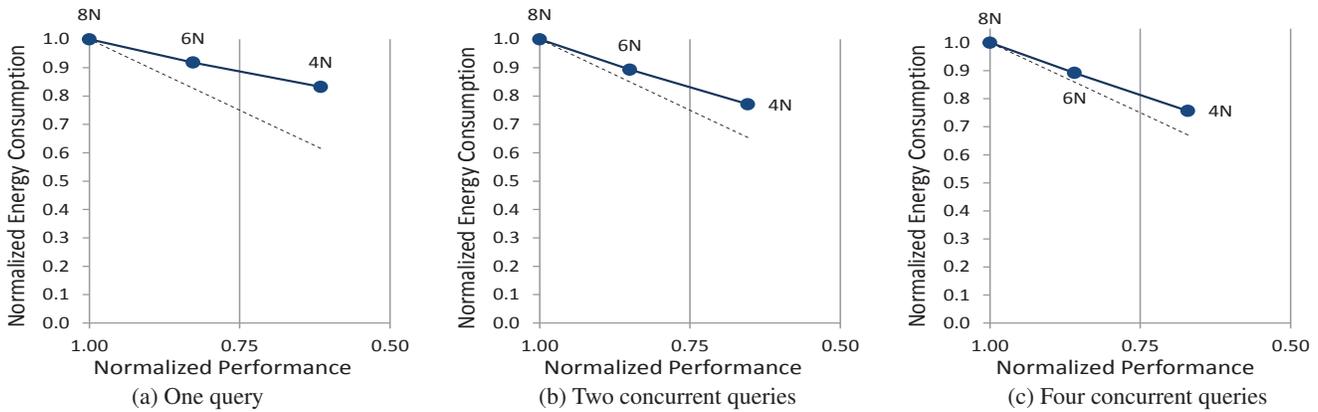

Figure 3: A partition incompatible TPC-H Q3 dual shuffle (exchange) hash join between LINEITEM and ORDERS (scale 1000) in P-store. Each subfigure (a-c) shows the energy consumption and performance ratios relative to the 8 node cluster reference point. The dotted line indicates the points where the EDP metric is constant.

Since our focus is on analytic SQL queries, we examine core parallel database operators, such as scans, joins, and exchanges. For these kinds of queries and operators we identify three categories of bottlenecks that can lead to underutilized hardware components in a cluster environment.

**Hardware bottleneck (network and disk):** Decision support queries often require repartitioning during query execution. Replication of tables on different partition keys can alleviate the need for repartitioning in some cases, but this option is limited to only small tables as it is often too expensive to replicate large tables.

The repartitioning step is often gated by the speed of the network interconnect, causing the processor and other resources of a node to idle while the node waits for data from the network. Additionally, an increase in network traffic on the cluster switches causes interference and further delays in communication. Future advances in networking technology are expected to be accompanied by advances in CPU capabilities, making this performance gap a persistent problem [30].

The balance between the network and disk subsystems can be easily disturbed with varying predicate selectivities which diminish the rate at which the storage subsystem can deliver qualified tuples to the network. As a result, underutilized CPU cores waiting to process operators at the higher levels of a query plan waste energy.

**Algorithmic bottleneck (broadcast):** For certain joins, the cheapest execution plan may be to broadcast a copy of the inner table (once all local predicates have been applied) to all the nodes so that the join is performed without re-partitioning the (potentially larger) outer table. Such a broadcast generally takes the same time to complete regardless of the number of participating nodes (e.g., for $m$GB of qualifying tuples and say 16 nodes, each node needs to receive $(15m/16)$GB of the data, while for 32 nodes this changes by a small amount to $(31m/32)$GB). As a result, scaling out to more nodes does not speed up this first phase of a join and it reduces the cluster energy efficiency.

**Data skew:** Although partitioning tools try to avoid data skew, even a small skew, can cause an imbalance in the utilization of the cluster nodes, especially as the system scales. Thus, data skew can easily create cluster and server imbalances even in highly tuned configurations. While we recognize data skew as a bottleneck, we leave an investigation of this critical issue as a part of future work.

### 4.2 P-store: A Custom-built Parallel Engine

P-store is built on top of a block-iterator tuple-scan module and a storage engine [16], that has scan, project, and select operators. To this engine, we added network exchange and hash join operators.

Since energy consumption is a function of average power and time, our goal was to make our engine perform at levels comparable to commercial systems. Therefore, it was imperative that our exchange operator is able to transfer over the network at rates near the limits of the physical hardware, and our operators never materialize tuples while maximizing utilization through multi-threaded concurrency.

### 4.3 Experiments

The purpose of these P-store experiments is to stress our "workhorse" exchange operator when performing partition-incompatible hash joins. By stressing the exchange operator, we wanted to see how the hash join operation behaves at different points in the cluster design space with respect to performance and energy consumption.

Our hash join query is between the LINEITEM and the ORDERS tables of TPC-H at a scale factor of 1000, similar to the Vertica experiments (see Section 3.1). We used eight of the *cluster-V* nodes described in Table 1. By measuring the CPU utilization, the average cluster power was found using our empirically derived model from the cluster-V column of Table 1, Section 3.

To explore the bottleneck of *partition-incompatible* joins, we hash partitioned the LINEITEM and ORDERS tables on their L_SHIPDATE and O_CUSTKEY attributes respectively, and examined the join between these two tables that is necessary for TPC-H Query 3. The LINEITEM table is projected to the L_ORDERKEY, L_EXTENDEDPRICE, L_DISCOUNT, and L_SHIPDATE columns, while the ORDERS table is projected to the O_ORDERKEY, O_ORDERDATE, O_SHIPPRIORITY, and O_CUSTKEY columns. To simulate the benefit of a columnar storage manager, for both tables, these four column projections (20B) were stored as tuples in memory for the scan operator to read. We apply a 5% selectivity predicate on both the tables using a predicate on the O_CUSTKEY attribute for ORDERS and a predicate on the L_SHIPDATE attribute for LINEITEM, as is done in TPC-H Query 3.

To perform the partition-incompatible join in this TPC-H query (Query 3), the hash join operator needs to build a hash table on the ORDERS table and then probe using the LINEITEM table. There are two ways to do this: (i) repartition both tables on the ORDERKEY attribute – a *dual shuffle*, and (ii) *broadcast* the qualifying ORDERS tuples to all nodes. With these two join methods, we show the effects of network and disk bottlenecks as well as algorithmic bottlenecks that affect energy efficiency.

#### 4.3.1 Dual Shuffle

By doing a repartitioning of both tables, our P-store results shows behavior that is similar to that of Vertica running TPC-H Query 12

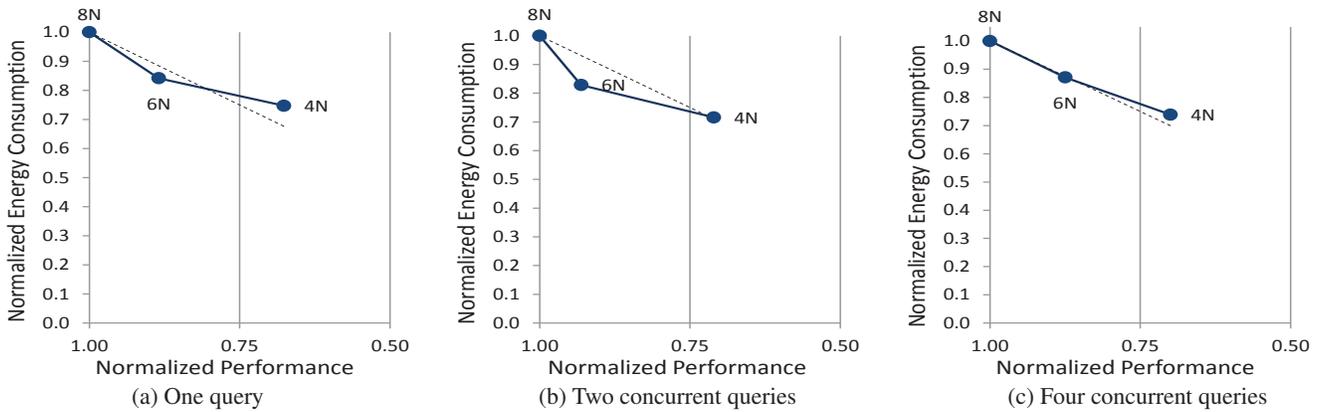

Figure 4: A partition incompatible TPC-H Q3 broadcast hash join between LINEITEM and ORDERS (scale 1000) in P-store. Each subfigure (a-c) shows the energy consumption and performance ratios relative to the 8 node cluster reference point. The dotted line indicates the points where the EDP metric is constant.

(Figure 1(a)). Using P-store, first the ORDERS table is repartitioned and the hash table is built on this table on-the-fly (no disk materialization) as tuples arrive over the network. After all the nodes have built their hash tables, the LINEITEM table is repartitioned and its tuples probe the hash tables on-the-fly.

In Figures 3(a)–(c), we see that poor performance scalability due to network bottlenecks can allow us to save energy by using fewer nodes. These figures show the comparison of relative energy consumption to relative performance of the hash join with 1, 2, and 4 independent concurrent joins being performed respectively on 4 to 8 node clusters. We increased the degree of concurrency to see how multiple simultaneous requests for the network resource affected the network bottleneck.

Our results show that 4 nodes (**4N**) always consumes less energy than 8 (**8N**). Also, as the concurrency level increases, the degree of energy savings also increases (i.e., the results move closer to the dashed EDP line). In Figure 3(a), halving the cluster size only results in a 38% decrease in performance and translates to almost 20% savings in energy consumption. At a concurrency level of two (Figure 3 (b)), the 4-node cluster has a 23% increase in energy savings over the 8-node cluster with a 35% penalty in performance. With 4 concurrent hash joins running on the cluster (Figure 3 (c)), the 4-node cluster saves 24% of the energy consumed by the 8-node cluster while losing 33% in performance.

The reason we see greater energy savings with more concurrent queries is because the CPU utilization does not proportionally increase with the increasing concurrency level. This behavior is due to the network being the bottleneck, and so the CPU stalls and idles as other hash joins also require the network resource.

To summarize, these results show that reducing the cluster size can save energy, but we pay for it with a proportionally greater loss in performance – i.e., these data points lie above the dashed EDP line. As we have mentioned previously, ideally we would prefer results that lie on or below this dotted curve. The next section shows that for broadcast joins, the trade-off is much more attractive.

### 4.3.2 Broadcast

The broadcast join method scans and filters the ORDERS table, and then sends the qualifying tuples to all the other nodes. Therefore, the full ORDERS hash table is built by each node and the LINEITEM table does not need to be repartitioned. To keep the full ORDERS hash table in memory, we increased the ORDERS table selectivity from 5% to 1% but held the LINEITEM table selectivity at 5%.

The results for this experiment are shown in Figure 4(a)–(c).

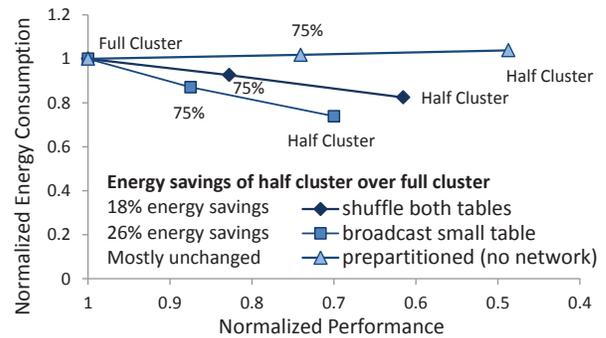

Figure 5: Network and algorithmic bottlenecks cause non-linear scalability in the partition-incompatible joins which mean smaller cluster designs consume less energy than larger designs. If the query is perfectly partitioned with ideal scalability (Vertica TPC-H Q1, Figure 2(a)), changing the number of nodes only affects performance and not energy consumption.

From these figures, we observe that the data points now lie on the EDP line, indicating that we proportionally trade an X% decrease in performance for an X% decrease in energy consumption compared to the 8 node case (**8N**). Similar to the dual shuffle cases, here we also used various concurrency levels – 1, 2, and 4.

In Figure 4(a), with one join running, the performance only decreases by 32% when we halve the cluster size from 8 to 4 nodes. With 2 and 4 concurrent joins, the performance decreases by around 30% in Figures 4(b,c) when halving the cluster size. This drop in performance causes the 4 node cluster to always save between 25% to 30% in energy consumption compared to the full 8 node cluster.

Compared to the dual shuffle join (see Figure 3), the broadcast join saves more energy when using 4 nodes rather than 8 nodes (data points lie closer to the EDP line in Figure 4). This means that the broadcast join suffers a higher degree of non-linear scalability than the dual shuffle join. This is because broadcasting the ORDERS table does not get faster with 8 nodes since every node must receive roughly the entire table (7/8) over the network (see Section 4.1).

## 4.4 Discussion

Figure 5 summarizes our findings. The energy consumption of the system, when running a 2-way join under different query execution plans, can vary significantly under different cluster configurations. Due to sub-linear performance speedup, when the join is not partition compatible ("shuffle both tables" and "broadcast small table"),

1689

| System | CPU (cores/threads) | RAM | Idle Power (W) |
|---|---|---|---|
| Workstation A | i7 920 (4/8) | 12GB | 93W |
| Workstation B | Xeon (4/4) | 24GB | 69W |
| Desktop | Atom (2/4) | 4GB | 28W |
| Laptop A | Core 2 Duo (2/2) | 4GB | 12W (screen off) |
| Laptop B | i7 620m (2/4) | 8GB | 11W (screen off) |

Table 2: Hardware configuration of different systems.

halving the cluster size does not result in doubling the response time, and this is one reason why the energy consumption at the "Half Cluster" configuration is lower than at "Full Cluster" point. Furthermore, we can see that when we use half of the cluster nodes instead of the full cluster, the broadcast join method saves more energy and suffers less performance penalty than the dual shuffle join. This is because, for the broadcast join approach, the hash join build phase does not get faster with more nodes and so 4 nodes is much more efficient than 8 nodes. We can put this result in perspective by comparing these results to the partitioned TPC-H Q1 result from Vertica, where the energy consumption is flat regardless of the cluster size.

## 5. ON CLUSTER DESIGN

Our empirical results using two off-the-shelf DBMSs and a custom parallel query execution kernel have shown that a number of key bottlenecks cause sub-linear speedup. These diminishing returns in performance, when additional cluster nodes are added, are a key reason why the energy consumption typically drops when we reduce the cluster size. Observing this reoccurring pattern of "smaller clusters can save energy" suggests that the presence of performance bottlenecks is a key factor for energy efficiency.

However, we have thus far ignored the other factor for total system energy consumption: namely, the power characteristics of a single node. As we demonstrate in this section, the hardware configuration (and performance capability) of each individual node also plays a significant role in improving the energy efficiency of parallel DBMSs, and needs to be considered for energy-efficient cluster design. By comparing some representative hardware configurations, we found one of our "Wimpy" laptop configurations yielded the lowest energy consumption in single node experiments. As such, we ask the question: *Given what we now know about performance bottlenecks and energy efficiency, what if we introduced these so-called "Wimpy" nodes into a parallel database cluster?*

In the remainder of this paper, we explore how query parameters, performance bottlenecks, and cluster design are the key factors that we should consider if we were to construct an energy-efficient cluster from scratch. We first show that Wimpy nodes consume less energy per query than "Beefy" nodes (Section 5.1). Then, using P-store we ran parallel hash joins on heterogeneous cluster designs (Section 5.2), and used these results to generate and validate a performance/energy consumption model (Sections 5.3). Finally, using this model, we explore the effects of some important query and cluster parameters (Section 5.4) to produce insights about the characteristics of the energy-efficient cluster design space (Section 5.5).

### 5.1 Energy Efficiency of Individual Nodes

In this section we demonstrate that non-server, low-power nodes consume significantly lower amounts of energy to perform the same task as a traditional "Beefy" server nodes. For the rest of the results in this paper, we had physical access to the nodes/servers so we measured power directly from the outlet using a WattsUp Pro meter. The meter provided a 1Hz sampling frequency with a measurement accuracy of +/- 1.5%. [5]

---
[5]We observed that the server CPU utilization/power models from these measurements validate our iLO2-based approach in Table 1.

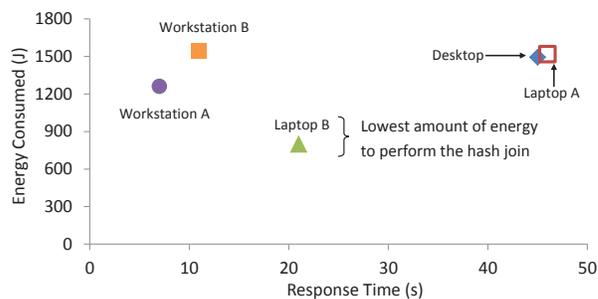

Figure 6: A hashjoin: 0.1M and 20M tuple (100B) tables.

We studied five systems with different power and performance characteristics, ranging from low-power, ultra-mobile laptops to high-end workstations. Laptops are optimized for power consumption and are typically equipped with mobile CPUs and SSDs, whereas high-end workstations are optimized for performance. The configuration and power consumption details are shown in Table 2.

To assess the energy efficiency of these systems, we used an in-memory database workload. This workload executes a basic hash join operation, and is designed to stress modern CPUs (the hash join code is cache-conscious and multi-threaded). Our hash join is between a 10MB table (100K cardinality, and 100 byte tuples) and a 2GB table (20M cardinality, and 100 byte tuples).

Figure 6 plots the energy consumed versus the hash join response time for the different configurations. From this figure, we observe that the Laptop B system consumes the lowest energy for processing this in-memory hash join. As expected, the high-end workstations exhibit the best performance (lowest response time). However, the workstations are not the best when we consider the energy consumption. The energy consumption of Laptop B is 800 Joules while the energy consumption of Workstation A is near 1300 Joules, even though it takes significantly longer to perform the in-memory join on the laptop. As this experiment suggests, low-power systems can reduce the average power that they draw more than they reduce performance, thereby reducing the energy consumption of running a database query. Since "Laptop B" consumed the least energy, we use it to represent a "Wimpy" server node when considering architectural designs of heterogeneous clusters.

Combining these results with the scalability and bottleneck observations (from Section 4.2), next we explore energy-efficient cluster design points, starting with experimental evidence of potential design opportunities.

### 5.2 Heterogeneous Clusters Design

We prototyped two different four node clusters and measured the energy efficiency for dual shuffle hash joins (see Section 4.3), using P-store. Each cluster node has a 1Gbs network card and is connected through a 10/100/1000 SMCGS5 switch. Node power was measured using the WattsUp power meters as described in Section 5.1. The cluster specifications are as follows.

**Beefy Cluster:** This cluster has four HP ProLiant SE326M1R2 servers with dual low-power quad-core Nehalem-class Xeon L5630 processors. Each node also has 32GB of memory and dual Crucial C300 256GB SSDs (only one was used for data storage). During our experiments, the average node power was 154W.

**2 Beefy/2 Wimpy:** This cluster has two "beefy" nodes from the Beefy cluster above, and two Laptop Bs (from Section 5.1), with i7 620m CPUs, 8GB of memory, and a Crucial C300 256GB SSD. During our experiments, the average laptop power was 37W.

Before we continue with the experimental results, there are two important notes to make. First, P-store does not support out-of-memory joins (2-pass joins), and therefore we either run in-memory joins across all the nodes (*homogeneous execution*), or we only perform the join on the nodes that have enough memory while the other nodes simply scan and filter data (*heterogeneous execution*). Since energy is a function of time and average power, in-memory join processing dramatically reduces the response time (as well as maintains high CPU utilization) and hence provides a substantial decrease in energy consumption per query. Second, our network has a peak capacity of 1Gbps, and therefore is typically the bottleneck for non-selective (where a high percentage of input tuples satisfy the selection predicates) queries. While a faster network could be used, as we discussed in Section 4.1, the network-CPU performance gap is likely to persist into the near future.

Using P-store, we ran the same hash join necessary for TPC-H Q3 between the LINEITEM and the ORDERS tables (as was done in the experiment described in Section 4.3), but with a scale factor of 400. The working sets (after projection) for the LINEITEM and the ORDERS tables are 48GB and 12GB respectively. We warmed our memory cache with as much of the working set as possible. The hash join is partition-incompatible on both the tables just as in Section 4.3 (i.e., dual shuffle is needed). We varied the selectivity on the LINEITEM table to 1%, 10%, 50%, and 100%. The ORDERS table had a selectivity of 1% and 10%. This gave us 8 different hash join workloads.

The Beefy cluster can run all of the 8 hash join workloads by first repartitioning ORDERS on the join key O_ORDERKEY, and building the hash table on each node as tuples arrive over the network. The LINEITEM table is then repartitioned on the L_ORDERKEY attribute, and each node probes their hash table on-the-fly as tuples arrive over the network. The beefy nodes have enough memory to cache the working set and build the in-memory hash table for the ORDERS table for both the 1% and 10% ORDERS selectivity values.

The hash join parameters cause the need for the 2 Beefy/2 Wimpy cluster to alternate between homogeneous and heterogeneous execution. The wimpy nodes only have 8GB of memory, and hence can only cache the 3GB ORDERS table partition and some of the 12GB LINEITEM table partition. Thus, for the 2 Beefy/2 Wimpy cluster, a 1% selectivity on ORDERS allows the hash table to fit in the laptop memory, and we can execute the join in the same way as the Beefy cluster – *homogeneous execution*. However, if the selectivity on ORDERS is low (i.e., the predicate matches ≥ 10% of the input tuples), then we can only leverage the wimpy nodes to scan and filter their partitions, and have them send all the qualifying data to the beefy nodes where the actual hash tables are built – *heterogeneous execution*. Similarly, the wimpy nodes scan and filter the LINEITEM (probe) tuples and repartition the data to the beefy nodes.

*Therefore, in our study, we consider two "wimpy" aspects of our mobile nodes: (1) their lower power and performance due to low-end CPUs, and (2) their lower memory capacity which constrains the execution strategies for evaluating a query.*

### 5.2.1 Small Hash Tables – Homogeneous Execution

Homogeneous parallel execution of a hash join requires a highly selective predicate that produces a small ORDERS hash table that can be stored in memory on all the nodes. We measured the response time (seconds) and the energy consumed (Joules) of both our cluster designs when running the hash join where the selectivity of the predicate on the ORDERS table was 1%. Figure 7 (a) shows the comparison of the all Beefy (**AB**) cluster to the 2 Beefy/2 Wimpy (**BW**) cluster when we vary the LINEITEM table selectivity (**L1,L10,L50,L100**) for a 1% selectivity on the ORDERS table (**O1**). We notice that for the 1% and 10% LINEITEM selectivity cases (the circle and square markers respectively), the **AB** cluster consumed less energy than the **BW**

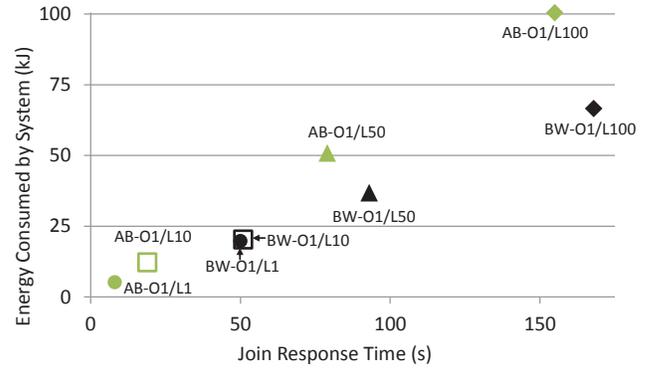

(a) all nodes build hash tables

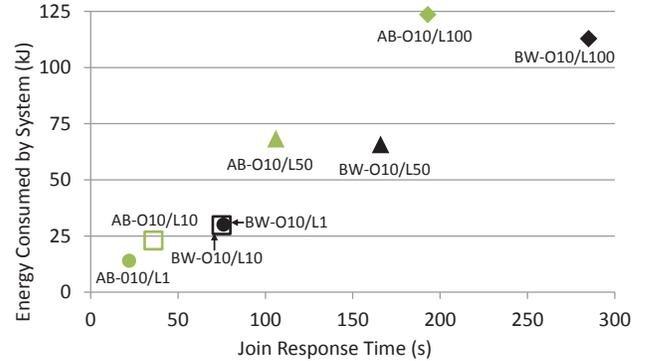

(b) beefy nodes build hash tables, wimpy nodes scan/filter data

Figure 7: P-store dual-shuffle hash join between LINEITEM (**L**) and ORDERS (**O**) tables at various selectivities (e.g., O10 means 10% selectivity on the ORDERS table). Empirical hash join energy efficiency of the all Beefy (**AB**) versus the 2 Beefy/2 Wimpy clusters (**BW**).

cluster. However, for the 50% LINEITEM selectivity case (triangle), the **BW** cluster saves 43% of the energy consumed over the **AB** cluster. When the LINEITEM table has no predicate (diamond), the **BW** cluster saves 56% of the energy consumed by the **AB** cluster.

Here we illustrate one of the bottlenecks we discussed above in Section 4: namely the network/disk. With 100% LINEITEM selectivity, the bottleneck is network bandwidth: the bottleneck response time of the **AB** cluster is 155s, while the response time of the **BW** cluster is 168s. With 1% LINEITEM selectivity, we are bound by the scan/filter limits of the wimpy, mobile nodes: the **AB** cluster finishes executing this join in 8s while the **BW** cluster takes 50s.

### 5.2.2 Large Hash Tables – Heterogeneous Execution

As mentioned above, in the cases where the hash tables are larger than the available memory in the wimpy, mobile nodes, the wimpy nodes of the mixed node cluster simply scan and filter the data for the Beefy nodes. Figure 7 (b) shows the comparison of the **AB** cluster to the **BW** cluster in terms of energy efficiency, similar to Figure 7 (a). Like our previous result for the 1% ORDERS selectivity case, the **BW** cluster consumes less energy than the **AB** cluster at a low LINEITEM selectivity (50% and 100%). The **BW** cluster consumes 7% and 13% less energy than the **AB** cluster at 50% and 100% LINEITEM selectivity respectively.

Thus, we see that in both situations, a heterogeneous cluster can offer improved energy efficiency at reduced performance. To further explore the full spectrum of available cluster and workload parameters that are important for cluster design, we built a model of P-store's performance and energy consumption. Our model is validated against these results in Figure 7. This model enables us to freely explore the cluster and query parameters that are important for cluster design.

| | | | |
|---|---|---|---|
| $T_{bld}$ | Build phase time (s) | $T_{prb}$ | Probe phase time (s) |
| $E_{bld}$ | Build phase energy (J) | $E_{prb}$ | Probe phase energy (J) |
| $N_B$ | # Beefy nodes | $N_W$ | # Wimpy nodes |
| $M_B$ | Beefy memory size (MB) | $M_W$ | Wimpy memory size (MB) |
| $I$ | Disk bandwidth (MB/s) | $L$ | Network bandwidth (MB/s) |
| $Bld$ | Hash join build table size (MB) | $Prb$ | Hash join probe table size (MB) |
| $S_{bld}$ | Build table predicate selectivity | $S_{prb}$ | Probe table predicate selectivity |

| | |
|---|---|
| $R_{Wbld}$ | Rate at which a Wimpy node builds its hash table (MB/s) |
| $R_{Bbld}$ | Rate at which a Beefy node builds its hash table (MB/s) |
| $U_{Wbld}$ | Wimpy node CPU bandwidth during the build phase |
| $U_{Bbld}$ | Beefy node CPU bandwidth during the build phase |
| $R_{Wprb}$ | Rate at which the Wimpy node probes its hash table (MB/s) |
| $R_{Bprb}$ | Rate at which the Beefy node probes its hash table (MB/s) |
| $U_{Wprb}$ | Wimpy node CPU bandwidth during the probe phase |
| $U_{Bprb}$ | Beefy node CPU bandwidth during the probe phase |
| $C_B = 5037$ | Maximum CPU bandwidth of a Beefy node (MB/s) |
| $C_W = 1129$ | Maximum CPU bandwidth of a Wimpy node (MB/s) |
| $G_B = 0.25$ | Beefy CPU utilization constants for P-store |
| $G_W = 0.13$ | Wimpy CPU utilization constants for P-store |
| $f_B(c) = 130.03 \times (100c)^{0.2369}$ (c=CPU util.) | Beefy node power model |
| $f_W(c) = 10.994 \times (100c)^{0.2875}$ (c=CPU util.) | Wimpy node power model |
| $H = M_W \geq (Bld * Bld_{sel})/(N_B + N_W)$ | Wimpy can build the hash table |

**Table 3: List of Model Variables**

## 5.3 Modeling P-store and Bottlenecks

Our model of P-store's performance and energy consumption behavior is aimed at understanding the nature of query parameters and scalability bottlenecks affecting performance and system energy consumption. The model predicts the performance and energy consumption of various different ways to execute a hash join. The input parameters that we consider are listed in Table 3.

From Table 3, note that our model makes a few simplifying assumptions. First, the disk configuration for both the Wimpy nodes and the Beefy nodes are the same and have the same bandwidth. Second, the same uniformity assumption has been made about the network capability of both node types. These assumptions matched our hardware setup, but we can easily extend our model to account for separate Wimpy and Beefy I/O bandwidths.

First, let us look at when the Wimpy nodes can build hash tables because they have enough memory to hold the hash tables (i.e., we do not have to run a 2-pass hash join). This is the case when $H$ is true (see Table 3).

**Homogeneous Execution:** In this case, all the nodes execute the same operator tree. We can divide the hash join into the build phase and the probe phase. During the build phase, we are either bound by (1) the effects of the disk bandwidth and the selectivity on the build table ; or (2) the network bandwidth:

$$R_{Bbld} = R_{Wbld} = \begin{cases} IS_{bld} & \text{if } IS_{bld} < L \\ \frac{(N_B + N_W)L}{(N_B + N_W - 1)} & \text{otherwise} \end{cases}$$

These two variables $R_{Bbld}$ and $R_{Wbld}$ give us the rates at which the build phase is proceeding (in MB/s). We also need to model the build phase CPU utilization to determine the power drawn by each node type. This is done by determining the amount of data that the Beefy CPU and the Wimpy CPU is processing per second, $U_{Bbld}$ and $U_{Wbld}$, and then dividing each of these values by the maximum measured rates for the CPU, $C_B$ and $C_W$ respectively. Finally, we need to add CPU constants that are inherent to P-store, which are $E_B$ and $E_W$ for the Beefy and Wimpy nodes respectively. This CPU utilization is the input for our server power functions, $f_B$ and $f_W$, for the Beefy and Wimpy nodes respectively (see Table 3).

$$U_{Bbld} = U_{Wbld} = \begin{cases} I & \text{if } IS_{bld} < L \\ \frac{(N_B + N_W)L}{(N_B + N_W - 1)} \div S_{bld} & \text{otherwise} \end{cases}$$

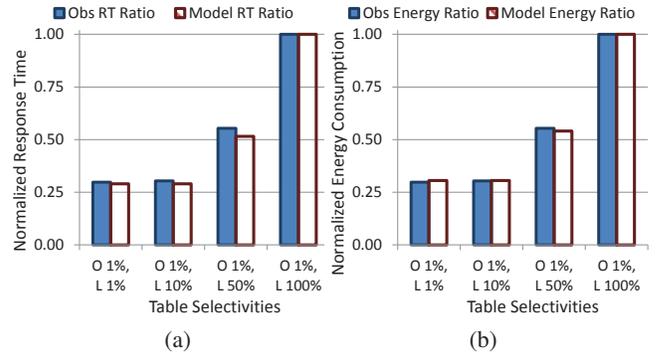

Figure 8: Performance and Energy model validation for the 2 Beefy/2 Wimpy case with 1% ORDERS selectivity and varying LINEITEM selectivity against observed data from Figure 7(a)

Therefore, we can now calculate the build phase response time and the build phase cluster energy consumption.

$$T_{bld} = \frac{Bld \times S_{bld}}{(N_B R_{Bbld}) + (N_W R_{Wbld})}$$

$$E_{bld} = T_{bld} \times (N_B f_B(G_B + \frac{U_{Bbld}}{C_B}) + N_W f_W(G_W + \frac{U_{Wbld}}{C_W}))$$

Since this is homogeneous execution, the probe phase can be modeled in the same way as the build phase but now using the probe table variables.

$$R_{Bprb} = R_{Wprb} = \begin{cases} IS_{prb} & \text{if } IS_{prb} < L \\ \frac{(N_B + N_W)L}{(N_B + N_W - 1)} & \text{otherwise} \end{cases}$$

And similarly, the CPU bandwidth during the probe phase is:

$$U_{Bprb} = U_{Wprb} = \begin{cases} I & \text{if } IS_{prb} < L \\ \frac{(N_B + N_W)L}{(N_B + N_W - 1)} \div S_{prb} & \text{otherwise} \end{cases}$$

Like the build phase, we can now calculate the probe phase response time and the probe phase cluster energy consumption.

$$T_{prb} = \frac{Prb \times S_{prb}}{(N_B R_{Bprb}) + (N_W R_{Wprb})}$$

$$E_{prb} = T_{prb} \times (N_B f_B(G_B + \frac{U_{Bprb}}{C_B}) + N_W f_W(G_W + \frac{U_{Wprb}}{C_W}))$$

With these two phases modeled, the total response time is simply $T_{bld} + T_{prb}$, and the total energy consumed is $E_{bld} + E_{prb}$.

**Heterogeneous Execution:** When the Wimpy nodes can not store the hash join hash table in memory ($H$ is false), P-store uses the Wimpy nodes as scan and filter nodes and only the Beefy nodes build the hash tables. Our model accounts for this by calculating the rates at which predicate-passing tuples are delivered to the Beefy nodes. In the same spirit as the above model for the homogeneous execution, the key factors are whether or not we are disk bound or network bound. However, since a smaller set of Beefy nodes need to receive the data from the entire cluster, in addition to out-bound network limitations from nodes sending data, there is an ingestion network limitation at the Beefy nodes, which becomes a performance bottleneck first. That is, the Beefy nodes that are building the hash tables can only receive data at the network's capacity even though there may be many Wimpy nodes trying to send data to them at a higher rate. For heterogeneous execution, we take this problem into account in our model. In the interest of space, we omit this model from this paper.

### 5.3.1 Model Validation

Next, we present validation of our P-store model using the real observed data of the 2 Beefy/2 Wimpy cluster in Section 5.2. Since the results in Section 5.2 were from warm-cache (some of the table data resides in memory and disk I/O may be necessary) hash joins, we changed the input parameters of our model to account for such behavior. For example, we changed the scan rate of the build phase to that of the maximum CPU bandwidth $C_W$ and $C_B$. In this way, the time it takes to finish the build phase is equal to the time it takes the CPU to process the build table at maximum speed plus the time to send the filter qualifying tuples over the network.

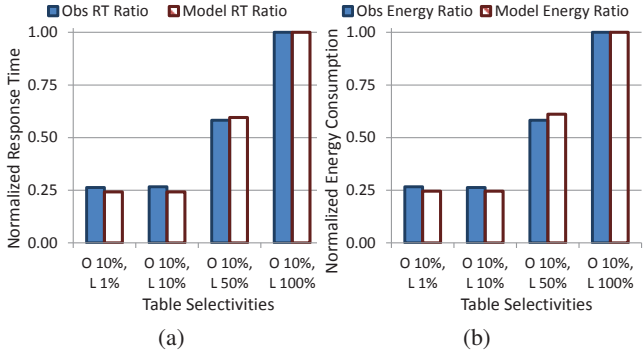
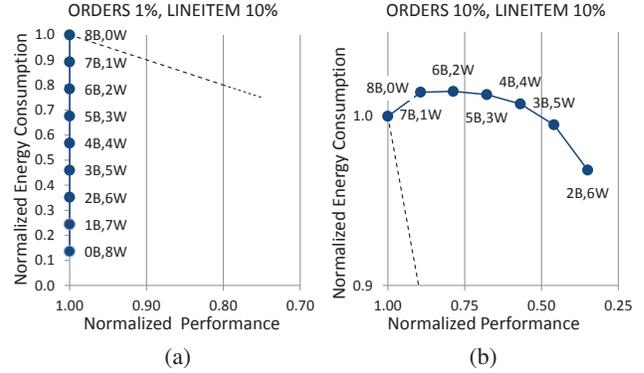

Figure 9: Performance and Energy model validation for the 2 Beefy/2 Wimpy case with 10% ORDERS selectivity and varying LINEITEM selectivity against observed data from Figure 7(b)

Figure 10: Modeled P-store dual-shuffle hash join performance and energy efficiency of an 8 node cluster made of various <u>B</u>eefy and <u>W</u>impy nodes. (a) Parallel execution is homogeneous: the Wimpy and Beefy nodes can build in-memory hash tables. (b) Parallel execution is heterogeneous: Wimpy nodes scan and filter their local data before shuffling it to the Beefy nodes for further processing. A constant EDP relative to the all Beefy cluster is shown by the dotted line.

For our model, we used the following hardware parameter settings: $M_B = 31000$, $M_W = 7000$, $N_B = N_W = 2$, $I = 270$, $L = 95$. Since our experiments used a different Beefy node (based on 2 L5630 Xeon CPUs) than the cluster-V nodes, the $f_B$ function for node power is given by $79.006 \times (100 * u)^{0.2451}$ and $C_B = 4034$.

Our validation results of our model against the 2 Beefy/2 Wimpy cluster results from Section 5.2 are shown in Figures 8 and 9. In Figure 8(a) and (b), we validate the response time and energy consumption results of our model against the 1% ORDERS table selectivity joins presented in Figure 7(a). The execution plans for the 1% ORDERS selectivity were homogeneous across all the nodes. In Figure 8, our model provided relative response time behavior and relative energy consumption behavior to the 100% LINEITEM selectivity result within 5% error compared to the observed data. Similarly, in Figure 9(a) and (b), we validated the relative response time and energy consumption results from our model for the 10% ORDERS selectivity joins from Figure 7(b). Here the error rate was within 10% compared to the observed data. The execution plans here were heterogeneous, with the Wimpy nodes only scanning and filtering data for the Beefy nodes.

With our model validated, we now explore a wider range of cluster designs beyond four nodes while also varying the query parameters, such as predicate selectivity. Our results reveal interesting opportunities for energy-efficient cluster design.

## 5.4 Exploring Query and Cluster Parameters

In this section, we explore what happens to query performance and energy consumption as we change the cluster design (the ratio of Beefy to Wimpy nodes) and query characteristics (the selectivity of build and probe tables), while executing the hash join query on P-store. This hash join is between a 700GB TPC-H ORDERS table and 2.8TB TPC-H LINEITEM table. We join these tables on their join attribute ORDERKEY. For our model, we used the following hardware parameter settings: $M_B = 47000$, $M_W = 7000$, $I = 1200$, $L = 100$. The memory settings correspond to those of our cluster-V nodes (Table 1, Section 3) and Laptop B (Table 2). We model the IO subsystem of our nodes as if they each had four of the Crucial 256GB SSDs that we used in Section 5.2, as well as the 1Gbps network interconnect of that experiment. We kept the remaining CPU parameters the same as those listed in Table 3.

We have already presented one of these results in Section 1. In Figure 1(b), we showed that increasing the ratio of Wimpy to Beefy nodes results in more energy-efficient configurations compared to the homogeneous cluster design consisting of only beefy nodes. In that figure, we used the model that we described in Section 5.3 to compare the energy consumption vs. performance trade-off for the hash join between the TPC-H ORDERS table and the LINEITEM table at 10% and 1% selectivity respectively.

In Figure 10(a), we show that the best cluster design point, when executing a hash join between an ORDERS table with 1% selectivity and a LINEITEM table with 10% selectivity, is to use all Wimpy nodes. In this case, since a 1% selectivity on the hash join build table means that each node only needs at least 875MB of memory to build in-memory hash tables, the parallel execution across all nodes is homogeneous. Since we have modeled the Wimpy and the Beefy nodes to have the same IO and network capabilities, we should not expect any performance degradation as we replace Beefy nodes with Wimpy nodes (i.e., the network and disk bottlenecks mask the performance limitations of the Wimpy nodes). This is shown in Figure 10(a) as the performance ratio remains 1.0 throughout all the configurations. Consequently, since performance does not degrade, we see that the energy consumed by the hash join drops by almost 90% because a Wimpy node power footprint is almost 10% of the Beefy node power footprint. In the interest of space, we omit the similar results for 1% ORDERS and 1% LINEITEM selectivity.

In Figure 10(b), the best cluster configuration for a hash join when the ORDERS table has the same 10% selectivity as in Figure 1(b) and the LINEITEM table also has 10% selectivity is to use all Beefy nodes. The Wimpy nodes do not have the 8.8GB of main memory that is needed to build the in-memory hash table, so we model them to scan and filter data for the Beefy nodes; i.e., we have a heterogeneous parallel execution. We varied the cluster design from the all Beefy to 2 Beefy and 6 Wimpy cases, after which, the aggregate Beefy memory cannot store the in-memory hash table. Here the results stand in stark contrast to the results shown in Figure 10(a), because there is not a significant energy savings when we use Wimpy nodes in the cluster. The reason is because with a 10% selectivity predicate on the tables, the server's IO subsystem has enough bandwidth to saturate the network interconnect, and the network becomes the bottleneck. Specifically, as we decrease the number of Beefy nodes – they are responsible for building and probing the hash tables – each Beefy node becomes more network bottlenecked ingesting data from all the other nodes in the cluster. Consequently, we see that the performance degrades severely, while the energy consumption does not drop below 95% of the **8B,0W** cluster reference point.

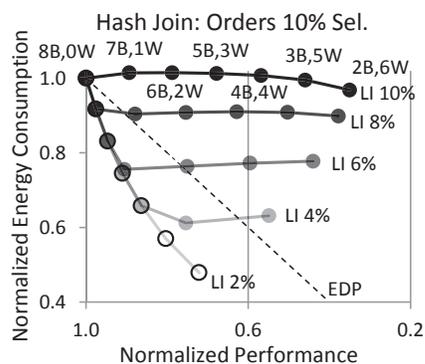

**Figure 11: Modeled P-store dual-shuffle hash join performance and energy efficiency of an 8 node cluster made of Beefy and Wimpy nodes. The join is between the TPC-H ORDERS table (10% sel.) and LINEITEM table (2-10% sel.). Parallel execution is heterogeneous across cluster nodes. A constant EDP relative to the all Beefy cluster is shown by the straight dotted line.**

This last result is interesting because we saw that for another heterogeneous execution plan in Figure 1(b), a 1% LINEITEM selectivity, saw significant wins in trading performance for energy savings. Naturally, we wanted to understand why we saw these different results when we only change the selectivities of the predicate on the LINEITEM table.

In Figure 11, we show that as we increase the selectivities of the predicate on the LINEITEM table from 10% to 2% (in 2% increments), given a 10% selectivity predicate on the ORDERS table, we begin to trade less performance for greater energy savings. Each curve represents a different LINEITEM selectivity, and as the curve moves to the right, each dot indicates more Wimpy nodes in the 8 node cluster. Again, we do not use fewer than 2 Beefy nodes because 1 Beefy node cannot build the entire hash table in memory. The dashed line indicates the constant EDP metric.

We notice that as we gradually decrease the number of LINEITEM tuples passing the selection filter (i.e., increase the LINEITEM table selectivity from 10% to 2%), the results start to trend downward below the EDP line. More interestingly, we notice that the results start to trend downward in a way such that the knee in the curves moves lower towards the cluster designs with more Wimpy nodes (right ends of the result curves). To the right of the knee, *the heterogeneous parallel plans saturate the Beefy node network ingestion during the probe phase*. To the left of the knee, the nodes delivering data to the Beefy nodes are sending data as fast as their IO subsystem (and table selectivity) can sustain. As the amount of probe (LINEITEM) data passing the selection filter decreases (the gradually lighter-shaded curves in Figure 11), the number of Wimpy nodes that are needed to saturate the inbound network port at the Beefy nodes increases, so the knee moves to the right end with more Wimpy nodes.

### 5.5 Summary

In this section, we have shown that there is an interesting interplay between the most energy-efficient cluster design (the ratio of Wimpy to Beefy nodes) and the query parameters such as predicate selectivity (Figure 11) for a simple parallel hash join query. Furthermore, heterogeneous cluster designs can trade proportionally less performance for greater energy savings (i.e., they lie below the EDP line) compared to a homogeneous cluster design.

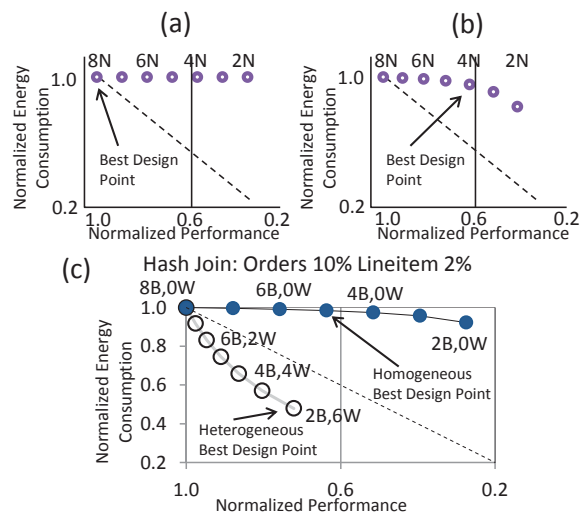

**Figure 12: Query scalability characterizations and energy-efficient design points when the target query performance is 40% of an eight Beefy node configuration, i.e., the target normalized performance is 0.6: (a) A highly scalable parallel workload: Use all available nodes, since the highest performing design point is also the most energy efficient. (b) A parallel workload with some scalability bottleneck (e.g., network I/O): Use the fewest nodes such that performance limits are still met. (c) Using fewer nodes reduces energy consumption on bottlenecked hash join but adding Wimpy nodes to supplement Beefy nodes can provide a more energy-efficient design point. The dashed EDP curve provides a reference for equal performance vs energy trade-offs.**

## 6. CLUSTER DESIGN PRINCIPLES

In this section we summarize guiding principles for building energy-efficient data processing clusters. Figure 12 outlines the summary of our findings, which we discuss in this section.

First, consider a query that is being run on a parallel data processing system. Using initial hardware calibration data and query optimizer information, consider the case when the query is deemed to be highly scalable. That is, the energy/performance model for the query looks like Figure 12(a). (We have seen empirical results like this, in Figures 2(a) and (b).) For such queries, the best cluster design point is to use the most resources (nodes) to finish the query as soon as possible. So, in Figure 12(a), we find that the largest cluster is the best design point.

However, if the query is not scalable (potentially due to bottlenecks), then the highest performing cluster design may not be the most energy-efficient design. (The results shown in Figures 1(a), 3, and 4 are examples of this case.) For such queries, reducing the cluster size decreases the query energy consumption, although at the cost of performance. In such situations, one should reduce the performance to meet any required target (e.g., performance targets specified in SLAs in cloud environments). Then, the system can reduce the server resource allocation accordingly. Thus, as we show in Figure 12(b), if the acceptable performance loss is 40%, then using 4 nodes is the best cluster design point for this query.

Finally, these previous design principles assume a homogeneous cluster. For queries that are not scalable, a heterogeneous cluster may provide an interesting design point. As before, in this case, to pick a good cluster design point, we start with an acceptable target performance. Then, heterogeneous cluster configurations may provide a better cluster design point compared to the best homogeneous cluster configuration. As an example, consider the P-store dual-shuffle hash join query in Figure 12(c), and an acceptable performance loss of 40% relative to an eight "Beefy" node homogeneous cluster design (labeled as 8B in the Figure). Here, the 5B configuration is the best homogeneous cluster design point. But if we substitute some of the Beefy nodes with lower-powered "Wimpy" nodes then with two Beefy nodes and six Wimpy nodes we consume less energy than the 5B case and also have better query performance. Notice that the heterogeneous design points are below the EDP curve, which means that in these designs one proportionally saved more energy than the proportional performance loss (compared to the 8B case). Thus, for non-scalable queries, a heterogeneous cluster configuration may provide a better design point, both from the energy efficiency and performance perspective, compared to homogeneous cluster designs.

Note, the work in this paper has focused on single queries. We acknowledge that to make these results more meaningful, we need to expand the study to include entire workloads, and to consider the overall cost of using only part of a cluster for a portion of the workload. In other words, we acknowledge that additional work is needed to produce a complete practical solution. But, we hope the insights in this paper provide a good starting point to seed future research in this area.

# 7. CONCLUSIONS AND FUTURE WORK

In this paper, we have studied the trade-offs between the performance and the energy consumption characteristics of analytical queries, for various cluster designs. We have found that the query scalability properties have a key impact in determining the interaction between the query performance and its energy consumption characteristics. We have summarized our findings (in Section 6) as initial guiding principles for building energy-efficient data processing clusters.

There are a number of directions for future work, including expanding this work to consider entire workloads, exploring heterogeneous execution plans to take advantage of heterogeneous clusters, examining the impact of data skew, and investigating the impact of dynamically varying multi-user workloads.

# 8. ACKNOWLEDGEMENTS

We thank David DeWitt, Jeffrey Naughton, Alan Halverson, Spyros Blanas, and Avrilia Floratou for their valuable input on this work. Part of this work was completed while Willis Lang, Stavros Harizopoulos, and Mehul Shah worked at HP Labs. This research was supported in part by a grant from the Microsoft Jim Gray Systems Lab and by the National Science Foundation under grant IIS-0963993.